\documentclass[12pt,a4paper,draftcls,onecolumn]{IEEEtran}
\usepackage{amsmath}
\usepackage{amssymb}
\usepackage{amsfonts}
\usepackage{amscd}
\usepackage{mathrsfs}
\usepackage[final]{graphicx}
\usepackage{color}
\usepackage{url}
\usepackage{verbatim}

\newtheorem{rem}{Remark}

\begin{document}

\title{Leveraging Coherent Distributed Space-Time Codes for Noncoherent Communication in Relay Networks via Training}

\author{G. Susinder Rajan and B. Sundar Rajan
\thanks{G. Susinder Rajan and B. Sundar Rajan are with the Department of Electrical Communication Engineering, Indian Institute of Science, Bangalore-560012, India. Email:\{susinder,bsrajan\}@ece.iisc.ernet.in.}}

%

\maketitle

\begin{abstract}

For point to point multiple input multiple output systems, Dayal-Brehler-Varanasi have proved that training codes achieve the same diversity order as that of the underlying coherent space time block code (STBC) if a simple minimum mean squared error estimate of the channel formed using the training part is employed for coherent detection of the underlying STBC. In this letter, a similar strategy involving a combination of training, channel estimation and detection in conjunction with existing coherent distributed STBCs is proposed for noncoherent communication in AF relay networks. Simulation results show that the proposed simple strategy outperforms distributed differential space-time coding for AF relay networks. Finally, the proposed strategy is extended to asynchronous relay networks using orthogonal frequency division multiplexing.
\end{abstract}
\section{Introduction}
\label{sec1}

Recently the idea of space time coding has been applied in wireless relay networks in the name of distributed space time coding to extract similar benefit as in point to point multiple input multiple output (MIMO) systems. Mainly there are two types of distributed space time coding techniques discussed in the literature: (i) decode and forward (DF) based distributed space time coding \cite{LaW}, wherein a subset (chosen based on some criteria) of the relay nodes decode the symbols from the source and transmit a row/column\footnote{Whether the relay transmits a column of a STBC or a row of a STBC depends on the
system model.} of a distributed space time block code (STBC) and (ii) amplify and forward (AF) based distributed space time coding \cite{JiH}, where all the relay nodes perform linear processing on the received symbols according to a distributed space time block code (DSTBC) and transmit the resulting symbols to the destination. AF based distributed space time coding is of special interest because the operations at the relay nodes are greatly simplified and moreover there is no need for every relay node to inform the destination once every quasi-static duration whether it will be participating in the distributed space time coding process  as is the case in DF based distributed space time coding \cite{LaW}. However, in \cite{JiH}, the destination was assumed to have perfect knowledge of all the channel fading gains from the source to the relays and those from the relays to the destination. To overcome the need for channel knowledge, distributed differential space time coding was studied in \cite{KiR,JiJ,OgH,RaR2}, which is essentially an extension of differential unitary space time coding for point to point MIMO systems to the relay network case. But distributed differential space time block code (DDSTBC) design is difficult compared to coherent DSTBC design because of the extra stringent conditions (we refer readers to \cite{JiJ,RaR2} for exact conditions) that need to be met by the codes. Moreover, all the codes in \cite{KiR,JiJ,OgH} for more than two relays have exponential encoding complexity. On the other hand, coherent DSTBCs with reduced maximum likelihood (ML) decoding complexity are available in \cite{RaR_IT,KiR2,MaH}.

Interestingly in \cite{DBV}, it was proved that for point to point MIMO systems, training codes\footnote{Each codeword of
a training code consists of a part known to the receiver (pilot) and a part that contains codeword(s) of a STBC designed for the coherent channel (in which receiver has perfect knowledge of the channel)} achieve the same diversity order as that of the underlying coherent STBC. This was shown to be possible if a simple minimum mean squared error (MMSE) estimate of the channel formed using the training part of the code is employed for coherent detection of the underlying STBC. The contributions of this letter are summarized as follows.
\begin{itemize}
\item Motivated by the results of \cite{DBV}, a similar training and channel estimation scheme is proposed to be used in conjunction with coherent distributed space time coding in AF relay networks as described in \cite{JiH}. An interesting feature of the proposed training scheme is that the relay nodes do not perform any channel estimation using the training symbols transmitted by the source but instead simply amplify and forward the received training symbols. The proposed strategy is shown to outperform the best known DDSTBCs \cite{KiR,JiJ,OgH,RaR2} using simulations. Also, it is shown that appropriate power allocation among the training and data symbols can further improve the error performance marginally.
\item Finally, this training based strategy is extended to asynchronous relay networks with no knowledge of the timing errors using the recently proposed Orthogonal Frequency Division Multiplexing (OFDM) based distributed space time coding \cite{RaR_icc}.
\end{itemize}

The rest of this letter is organized as follows. The proposed training scheme along with channel estimation is described in Section \ref{sec2}. Extension to the asynchronous relay network case is addressed in Section \ref{sec3}. Simulation results comprise Section \ref{sec4} and a conclusions are presented in Section \ref{sec5}.

\textbf{Notation:} Vectors and matrices are represented by lowercase and uppercase boldface characters respectively. An identity matrix of size $N\times N$ will be denoted by $\mathbf{I_N}$. A complex Gaussian vector with zero mean and covariance matrix $\mathbf{\Omega}$ will be denoted by $\mathcal{CN}(0,\mathbf{\Omega})$.


\section{Proposed Training Based Strategy}
\label{sec2}

In this section, we briefly review the distributed space time coding protocol for AF relay networks in \cite{JiH}, make some crucial observations and then proceed to describe the proposed training based strategy.

Consider a wireless relay network consisting of a source node, a destination node and $R$ relay nodes $U_1,U_2,\dots,U_R$ which aid the source in communicating information to the destination. All the nodes are assumed to be equipped with a half duplex constrained, single antenna transceiver. The wireless channels between the terminals are assumed to be quasi-static and flat fading. The channel fading gains from the source to the $i$-th relay, $f_i$ and those from the $i$-th relay to the destination $g_i$ are all assumed to be independent and identically distributed (i.i.d) complex Gaussian random variables with zero mean and unit variance. Symbol synchronization and carrier frequency synchronization are assumed among all the nodes.

\subsection{Observations from Coherent Distributed Space Time Coding}
\label{subsec2_1}

In order to explain coherent distributed space time coding, we shall assume in this subsection alone that the destination has perfect knowledge of all the channel fading gains $f_i,g_i,~i=1,\dots,R$. Every transmission cycle from the source to the destination is comprised of two phases. In the first phase, the source transmits a vector $\mathbf{z}=\left[\begin{array}{cccc} z_1 & z_2 & \dots & z_{T_1}\end{array}\right]^T$ composed of $T_1$ complex symbols $z_i,~i=1,\dots,T_1$ to all the $R$ relays using a fraction $\pi_1$ of the total power $P_d$ for data transmission. The vector $\mathbf{z}$ satisfies $\mathrm{E}[\mathbf{z}^H\mathbf{z}]=T_1$ and $P_d$ denotes the total average power spent by the source and the relays for communicating data to the destination. The received vector at the $i$-th relay is then given by $\mathbf{r}_i=\sqrt{\pi_1P_d}f_i\mathbf{z}+\mathbf{v}_i$ where, $\mathbf{v}_i\sim\mathcal{CN}(0,\mathbf{I_{T_1}})$ represents the additive noise at the $i$-th relay.

In the second phase, the $i$-th relay transmits $\mathbf{t}_i=\sqrt{\frac{\pi_2P_d}{\pi_1P_d+1}}\mathbf{B}_i\mathbf{r}_i$ or $\mathbf{t}_i=\sqrt{\frac{\pi_2P_d}{\pi_1P_d+1}}\mathbf{B}_i\mathbf{r}_i^*$ to the destination, where $\mathbf{B}_i\in\mathbb{C}^{T_2\times T_1}$ is called the `relay matrix'. Without loss of generality we may assume that the first $M$ relays linearly process $\mathbf{r_i}$ and the remaining $R-M$ relays linearly process $\mathbf{r_i}^*$. Under the assumption that the quasi-static duration of the channel is much greater than $2R$ channel uses, the received vector at the destination can be expressed as $\mathbf{y}=\sum_{i=1}^{R}g_i\mathbf{t}_i+\mathbf{w}=\sqrt{\frac{\pi_1\pi_2P_d^2}{\pi_1P_d+1}}\mathbf{Xh}+\mathbf{n}$ where, $\mathbf{X}=\left[\begin{array}{ccccccc}\mathbf{B_1z} & \dots & \mathbf{B_Mz} & \mathbf{B_{M+1}z}^* & \dots & \mathbf{B_Rz}^*\end{array}\right]$,
\begin{equation}
\label{eqn_dstbc}
\mathbf{h}=\left[\begin{array}{ccccccc}f_1g_1 & f_2g_2 & \dots f_Mg_M & f_{M+1}^*g_{M+1} & \dots & f_R^*g_R\end{array}\right]^T,
\end{equation}
$\mathbf{n}=\sqrt{\frac{\pi_2P_d}{\pi_1P_d+1}}\left(\sum_{i=1}^{M}g_i\mathbf{B}_i\mathbf{v}_i+\sum_{i=M+1}^{R}g_i\mathbf{B}_i\mathbf{v}_i^*\right)+\mathbf{w}$
\noindent and $\mathbf{w}\sim\mathcal{CN}(0,\mathbf{I_{T_2}})$ represents the additive noise at the destination. The power allocation factors $\pi_1$ and $\pi_2$ are chosen to satisfy $\pi_1P_d+\pi_2P_dR=2P_d$. The covariance matrix of $\mathbf{n}$ is given by $\mathbf{\Gamma}=\mathrm{E}[\mathbf{n}\mathbf{n}^H]=\mathbf{I_{T_2}}+\frac{\pi_2P_d}{\pi_1P_d+1}(\sum_{i=1}^{R}|g_i|^2\mathbf{B}_i\mathbf{B}_i^H)$. Let the DSTBC $\mathscr{C}$ denote the set of all possible codeword matrices $\mathbf{X}$. Then the ML decoder is given by
\begin{equation}
\label{eqn_ml_dstbc}
\mathbf{\hat{X}}=\arg\min_{\mathbf{X}\in\mathscr{C}}\parallel\mathbf{\Gamma}^{-\frac{1}{2}}(\mathbf{y}-\sqrt{\frac{\pi_1\pi_2P_d^2}{\pi_1P_d+1}}\mathbf{X}\mathbf{h})\parallel_F^2.
\end{equation}

Note from \eqref{eqn_ml_dstbc} that the ML decoder in general requires the knowledge\footnote{$\Gamma$ requires knowledge of the $g_i$'s and $\mathbf{h}$ requires knowledge of $f_ig_i,~i=1,\dots,M$ and $f_i^*g_i,~i=M+1,\dots,R$ which together imply knowledge of $f_i,g_i,~i=1,\dots,R$.} of all the channel fading gains $f_i,g_i,~i=1,\dots,R$. Consider the following decoder:

\begin{equation}
\label{eqn_subml_dstbc}
\mathbf{\hat{X}}=\arg\min_{\mathbf{X}\in\mathscr{C}}\parallel\mathbf{y}-\sqrt{\frac{\pi_1\pi_2P_d^2}{\pi_1P_d+1}}\mathbf{X}\mathbf{h}\parallel_F^2.
\end{equation}

\begin{rem}
The decoder in \eqref{eqn_subml_dstbc} is suboptimal in general and coincides with the ML decoder for the case when $\mathbf{\Gamma}$ is a scaled identity matrix. The relay matrices for all the codes in \cite{JiH,KiR2,EOK,MaH} and some of the codes in \cite{RaR_IT} are unitary. For the case when $\mathbf{B}_i\mathbf{B}_i^H$ is a diagonal matrix for all $i=1,2,\dots,R$ ($\mathbf{\Gamma}$ is a diagonal matrix for this case), the performance of the suboptimal decoder in \eqref{eqn_subml_dstbc} differs from that of the ML decoder \eqref{eqn_ml_dstbc} only by coding gain and the diversity gain is retained. This can be proved on similar lines as in the proof of Theorem 7 in \cite{RaR_IT}. The class of DSTBCs from precoded co-ordinate interleaved orthogonal designs in \cite{RaR_IT} is an example for the case of diagonal $\mathbf{\Gamma}$ matrix.
\end{rem}

The decoder in \eqref{eqn_subml_dstbc} requires only the knowledge of $\mathbf{h}$ and not necessarily the knowledge of all the individual channel fading gains $f_i,g_i,~i=1,2,\dots,R$. The training strategy to be described in the sequel essentially exploits this crucial observation.

\subsection{Training cycle}
\label{subsec2_2}

Note from the previous subsection that one data transmission cycle comprises of $T_1+T_2$ channel uses. In the proposed training strategy, we introduce a training cycle comprising of $R+1$ channel uses for channel estimation before the start of data transmission cycle. We assume that the quasi-static duration of the channel is greater than $(R+1)+F(T_1+T_2)$ channel uses where, $F$ denotes the total number of data transmission cycles that can be accommodated within the channel quasi-static duration. Thus, for $F=1$, $T_1=T_2=R$, the minimum channel quasi-static duration required for the proposed strategy is $3R+1$ channel uses. Let $P_t$ be the total average power spent by the source and the relays during the training cycle. Thus, the total average power $P$ used by the source and the relays is $P=\frac{P_t(R+1)+P_dF(T_1+T_2)}{R+1+F(T_1+T_2)}$.

In the first phase of the training cycle, the source transmits the complex number $1$ to all the relays using a fraction $\pi_1$ of the total power $P_t$ dedicated for training. The received symbol at the $i$-th relay denoted by $\hat{r}_i$ is given by $\hat{r}_i=\sqrt{\pi_1 P_t}f_i+\hat{v}_i$ where $\hat{v}_i\sim\mathcal{CN}(0,1)$ is the additive noise at the $i$-th relay.

The second phase of the training cycle comprises of $R$ channel uses, out of which one channel use is assigned to every relay node. Without loss of generality, we may assume that the $i$-th time slot is assigned to the $i$-th relay. Furthermore, we assume that the value of $M$ to be used during the data transmission cycle is already decided. During its assigned time slot, the $i$-th relay transmits $\hat{t}_i=\left\{\begin{array}{cc}\sqrt{\frac{\pi_2P_tR}{\pi_1P_t+1}}\hat{r}_i=\sqrt{\frac{\pi_1\pi_2P_t^2R}{\pi_1P_t+1}}f_i+\sqrt{\frac{\pi_2P_tR}{\pi_1P_t+1}}\hat{v}_i, & \mathrm{if}~i\leq M\\
\sqrt{\frac{\pi_2P_tR}{\pi_1P_t+1}}\hat{r}_i^*=\sqrt{\frac{\pi_1\pi_2P_t^2R}{\pi_1P_t+1}}f_i^*+\sqrt{\frac{\pi_2P_tR}{\pi_1P_t+1}}\hat{v}_i^*, & \mathrm{if}~i>M
\end{array}\right.$. At the end of the training cycle, the received vector $\hat{y}$ at the destination is given as follows:
\begin{equation}
\label{eqn_training}
\mathbf{\hat{y}}=\sqrt{\frac{\pi_1\pi_2P_t^2R}{\pi_1P_t+1}}\mathbf{I_R}\mathbf{h}+\mathbf{\hat{n}}
\end{equation}
\noindent where, $\mathbf{\hat{n}}=\sqrt{\frac{\pi_2P_tR}{\pi_1P_t+1}}\left[\begin{array}{cccccc} g_1\hat{v}_1 & \dots & g_M\hat{v}_M & g_{M+1}\hat{v}_{M+1}^* & \dots & g_R\hat{v}_R^*\end{array}\right]^T+\mathbf{\hat{w}}$, $\mathbf{h}$ is same as that given in \eqref{eqn_dstbc} and $\mathbf{\hat{w}}\sim\mathcal{CN}(0,\mathbf{I_R})$ is the additive noise at the destination. The entire transmission from source to destination is illustrated pictorially in Fig. \ref{fig_training} and Fig. \ref{fig_data}.

Note that the entries of $\mathbf{h}$ as well as $\mathbf{\hat{n}}$ are not complex Gaussian distributed since they involve terms that are product of complex Gaussian random variables. To be precise, the entries of $\mathbf{h}$ are i.i.d random variables with mean $0$ and variance $1$. Similarly, the entries of $\mathbf{n_t}$ are   i.i.d random variables with mean $0$ and variance $(\frac{\pi_2P_tR}{\pi_1P_t+1}+1)$. For the point to point MIMO case, where the channel and additive receiver noise are modeled as complex Gaussian, Dayal-Brehler-Varanasi in \cite{DBV} have proposed a simple linear channel estimator. In this letter, we propose to employ a similar estimator for the equivalent channel $\mathbf{h}$ as follows:

\begin{equation}
\label{eqn_ch_estimator}
\hat{\mathbf{h}}=\sqrt{\frac{\pi_1\pi_2P_t^2R}{\pi_1P_t+1}}\left(\frac{\pi_2P_tR+\pi_1\pi_2P_t^2R}{\pi_1P_t+1}+1\right)^{-1}\mathbf{\hat{y}}
\end{equation}

Now using the estimate $\hat{\mathbf{h}}$, coherent DSTBC decoding can be done in every data transmission cycle, as $\mathbf{\hat{X}}=\arg\min_{\mathbf{X}\in\mathscr{C}}\parallel\mathbf{y}-\sqrt{\frac{\pi_1\pi_2P_d^2}{\pi_1P_d+1}}\mathbf{X}\hat{\mathbf{h}}\parallel_F^2$. Thus, coherent DSTBCs \cite{RaR_IT,KiR2,JiJ2,EOK,MaH} can be employed in noncoherent relay networks via the proposed training scheme. We would like to mention that there may be better channel estimation techniques than the one described by \eqref{eqn_ch_estimator}, but this is beyond the scope of this letter. However, the simulation results in section \ref{sec5} show that a simple channel estimator as in \eqref{eqn_ch_estimator} is good enough to outperform the best known DDSTBCs.

\section{Training Strategy For Asynchronous Relay Networks}
\label{sec3}

\begin{figure}[p]
\centering
\input{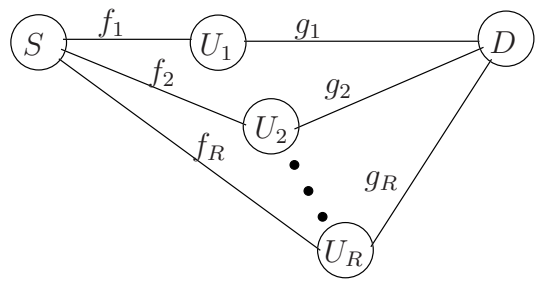}
\caption{Asynchronous wireless relay network}
\label{fig_network}
\end{figure}

The training strategy described in the previous section assumes that the transmissions from all the relays are symbol synchronous with reference to the destination. In this section, we relax this assumption and extend the proposed training strategy to asynchronous relay networks with no knowledge of the timing errors of the relay transmissions. However we shall assume that the maximum of the relative timing errors from the source to the destination is known. An asynchronous wireless relay network is depicted in Fig.\ref{fig_network}. Let $\tau_i$ denote the overall relative timing error of the signals arrived at the destination node from the $i$-th relay node. Without loss of generality, we assume that $\tau_1=0$, $\tau_{i+1}\geq \tau_i,i=1,\dots,R-1$.

Recently there have been several works \cite{RaR_icc,RaR_IT,GuX,ZLiX,YLiX,EK,EKJ} on distributed space time coding for asynchronous relay networks, some of which employ OFDM. The proposed scheme relies on the OFDM based distributed space time coding in \cite{RaR_icc,RaR_IT}, which is essentially distributed space time coding over OFDM symbols and the cyclic prefix (CP) of OFDM is used to mitigate the effects of symbol asynchronism. The number of sub-carriers $N$ and the length of the cyclic prefix (CP) $l_{cp}$ are chosen such that $l_{cp}\geq\max_{i=1,2,\dots,R}\left\{\tau_i\right\}$. The channel quasi-static duration assumed for this strategy is $\left((R+1)+F(2R)\right)\left(N+l_{cp}\right)$ channel uses.

As for the synchronous case, there will be a training cycle before the start of data transmission from the source. In the first phase of the training cycle, the source takes the $N$ point inverse discrete Fourier transform (IDFT) of the $N$ length vector $\mathbf{p}=\left[\begin{array}{cccc}1 & 1 & \dots & 1\end{array}\right]^T$ and adds a CP of length $l_{cp}$ to form a OFDM symbol $\mathbf{\bar{p}}$. This OFDM symbol is transmitted to the relays using a fraction $\pi_1$ of the total power $P_t$. The $i$-th relay receives $\mathbf{\hat{r}}_i=\sqrt{\pi_1 P_t}f_i\mathbf{\bar{p}}+\mathbf{\bar{\hat{v}}}_i$ where $\mathbf{\bar{\hat{v}}}_i\sim\mathcal{CN}(0,\mathbf{I_{N+l_{cp}}})$ is the additive noise at the $i$-th relay. The second phase of the training cycle comprises of $R$ OFDM time slots and the $i$-th relay is allotted the $i$-th OFDM time slot for transmission. During its scheduled time slot, the $i$-th relay transmits
$\mathbf{\hat{t}}_i=\left\{\begin{array}{cc}\sqrt{\frac{\pi_2RP_t}{\pi_1P_t+1}}\mathbf{\hat{r}}_i, &\mathrm{if}~i\leq M\\
\sqrt{\frac{\pi_2RP_t}{\pi_1P_t+1}}\zeta\left((\mathbf{\hat{r}}_i)^{*}\right), & \mathrm{if}~i>M \end{array}\right.$ where $\zeta(.)$ denotes the time reversal operation, i.e., $\zeta(\mathbf{r}(n))\triangleq \mathbf{r}(N+l_{cp}-n)$. The destination receives $R$ OFDM symbols which are processed as follows:
\begin{enumerate}
\item Remove the CP for the first $M$ OFDM symbols.
\item For the remaining OFDM symbols, remove CP to get a $N$-length vector. Then shift the last $l_{cp}$ samples of the $N$-length vector as the first $l_{cp}$ samples.
\end{enumerate}

Discrete Fourier transform (DFT) is then applied on the resulting $R$ vectors to obtain $\mathbf{\hat{x}}_j=\left[\begin{array}{cccc}\hat{y}_{0,j} & \hat{y}_{1,j} & \dots & \hat{y}_{N-1,j}\end{array}\right]^T,~j=1,2,\dots,R$. Let $\mathbf{\hat{w}}_j=\left[\begin{array}{cccc}\hat{w}_{0,j} & \hat{w}_{1,j} & \dots & \hat{w}_{N-1,j}\end{array}\right]^T$ represent the additive noise at the destination node in the $j$-th OFDM time slot and let $\mathbf{\hat{v}}_j=\left[\begin{array}{cccc}\hat{v}_{0,j} & \hat{v}_{1,j} & \dots & \hat{v}_{N-1,j}\end{array}\right]^T$ denote the DFT of $\mathbf{\bar{\hat{v}}}_j$ after CP removal. Note that a delay $\tau$ in the time domain translates to a phase change of $e^{-\frac{i2\pi k\tau}{N}}$ in the $k$-th sub carrier. Now using the identities $(\mathrm{DFT}(\mathbf{x}))^*=\mathrm{IDFT}(\mathbf{x}^*)$, $(\mathrm{IDFT}(\mathbf{x}))^*=\mathrm{DFT}(\mathbf{x}^*)$, $\mathrm{DFT}(\zeta(\mathrm{DFT}(\mathbf{x})))=\mathbf{x}$, $\mathbf{p}^*=\mathbf{p}$ we have in the $j$-th OFDM time slot
$$
\mathbf{\hat{x}}_j=\left\{\begin{array}{cc}f_jg_j\sqrt{\frac{\pi_1\pi_2RP_t^2}{\pi_1P_t+1}}\mathbf{p}\circ \mathbf{d^{\tau_j}}+\sqrt{\frac{\pi_2RP_t}{\pi_1P_t+1}}g_j\mathbf{\hat{v}}_j\circ \mathbf{d^{\tau_j}}+\mathbf{\hat{w}}_j & \mathrm{if}~j\leq M\\
f_j^*g_j\sqrt{\frac{\pi_1\pi_2RP_t^2}{\pi_1P_t+1}}\mathbf{p}\circ \mathbf{d^{\tau_j}}+\sqrt{\frac{\pi_2RP_t}{\pi_1P_t+1}}g_j\mathbf{\hat{v}}_j^*\circ \mathbf{d^{\tau_j}}+\mathbf{\hat{w}}_j & \mathrm{if}~j>M
\end{array}\right.
$$
\noindent where, $\mathbf{d^{\tau_j}}=\left[\begin{array}{cccc}1 & e^{-\frac{i2\pi\tau_j}{N}} & \dots & e^{-\frac{i2\pi\tau_j(N-1)}{N}} \end{array}\right]^T$ and $\circ$ denotes Hadamard product. Thus, in each sub-carrier $k,~0\leq k\leq N-1$, we get
\begin{equation}
\label{eqn_sys_model}
\mathbf{\hat{y}}_k=\left[\begin{array}{cccc}\hat{y}_{k,1} & \hat{y}_{k,2} & \dots & \hat{y}_{k,R}\end{array}\right]^T=\sqrt{\frac{\pi_1\pi_2RP_t^2}{\pi_1P_t+1}}\mathbf{I_R}\mathbf{h}_k+\mathbf{\hat{n}}_k
\end{equation}
\noindent where,
\begin{equation}
\label{eqn_channel}
\mathbf{h}_k=\left[\begin{array}{ccccccc} f_1g_1 & u_k^{\tau_2}f_2g_2 & \dots &
u_k^{\tau_{M}}f_{M}g_{M} &
u_k^{\tau_{M+1}}f_{M+1}^*g_{M+1} & \dots &
u_k^{\tau_R}f_R^*g_R\end{array}\right]^T,
\end{equation}
\noindent $u_k^{\tau_i}=e^{-\frac{i2\pi k\tau_i}{N}}$ and
$$
\begin{array}{rl}
\mathbf{\hat{n}}_k=&\sqrt{\frac{\pi_2P_tR}{\pi_1P_t+1}}\left[\begin{array}{cccccc}u_k^{\tau_1}g_1\hat{v}_{k,1} & \dots &
u_k^{\tau_{M}}g_{M}\hat{v}_{k,M} &
u_k^{\tau_{M+1}}g_{M+1}\hat{v}_{k,M+1}^* & \dots &
u_k^{\tau_R}g_{R}\hat{v}_{k,R}^*\end{array}\right]\\
&+\left[\begin{array}{cccc}\hat{w}_{k,1} & \hat{w}_{k,2} & \dots \hat{w}_{k,R}\end{array}\right]^T.
\end{array}
$$

Analogous to the synchronous case, we propose to estimate the equivalent channel matrix $\mathbf{h}_k$ from \eqref{eqn_sys_model} as $\mathbf{\hat{h}}_k=\sqrt{\frac{\pi_1\pi_2RP_t^2R}{\pi_1P_t+1}}\left(\frac{\pi_2P_tR+\pi_1\pi_2P_t^2R}{\pi_1P_t+1}+1\right)^{-1}\mathbf{\hat{y}}_k$. After the training cycle, the data transmission cycle starts for which refer the readers to \cite{RaR_icc} and section IV of \cite{RaR_IT} for a detailed explanation. In essence, a DSTBC is seen by the destination in every sub-carrier and the equivalent channel seen by the destination in the $k$-th sub-carrier is precisely the matrix $\mathbf{h}_k$, whose estimated value is available at the end of the training cycle. As for the synchronous case (see \eqref{eqn_subml_dstbc}), we propose to ignore the covariance matrix of the equivalent noise while performing data detection .

\section{Simulation Results}
\label{sec4}

In this section, simulations are used to compare the error performance of the proposed strategy against the best known DDSTBC for $2$ relays\cite{JiJ} and $4$ relays\cite{RaR2}. Note that for $4$ relays, the DDSTBCs in \cite{RaR2} were shown to outperform the codes reported in \cite{KiR,JiJ,OgH} in both complexity as well as performance. For all the simulations, we set $\pi_1=1$, $\pi_2=\frac{1}{R}$ (as suggested in \cite{JiH}), $T_1=T_2=4$ and $F=50$. The channel fading gains $f_i,g_i,~i=1,\dots,R$ are each generated independently following a complex Gaussian distribution with mean $0$ and unit variance\footnote{This is a suitable assumption for the case when the relays are approximately equidistant from both the source as well as the destination.}. The decoder used for the proposed scheme is the one described by \eqref{eqn_subml_dstbc} and for the DDSTBC case, the decoder proposed in \cite{RaR2} has been used. We chose $P_t=(1+\alpha)P_d$, where $\alpha$ denotes the power boost factor to allow for power boosting to the pilot symbols. In order to quantify the loss in error performance due to channel estimation errors in the proposed strategy, the performance of the corresponding coherent DSTBC (assuming perfect channel knowledge) is taken as the reference.

\begin{figure}[p]
\centering
\includegraphics[width=6.5 in]{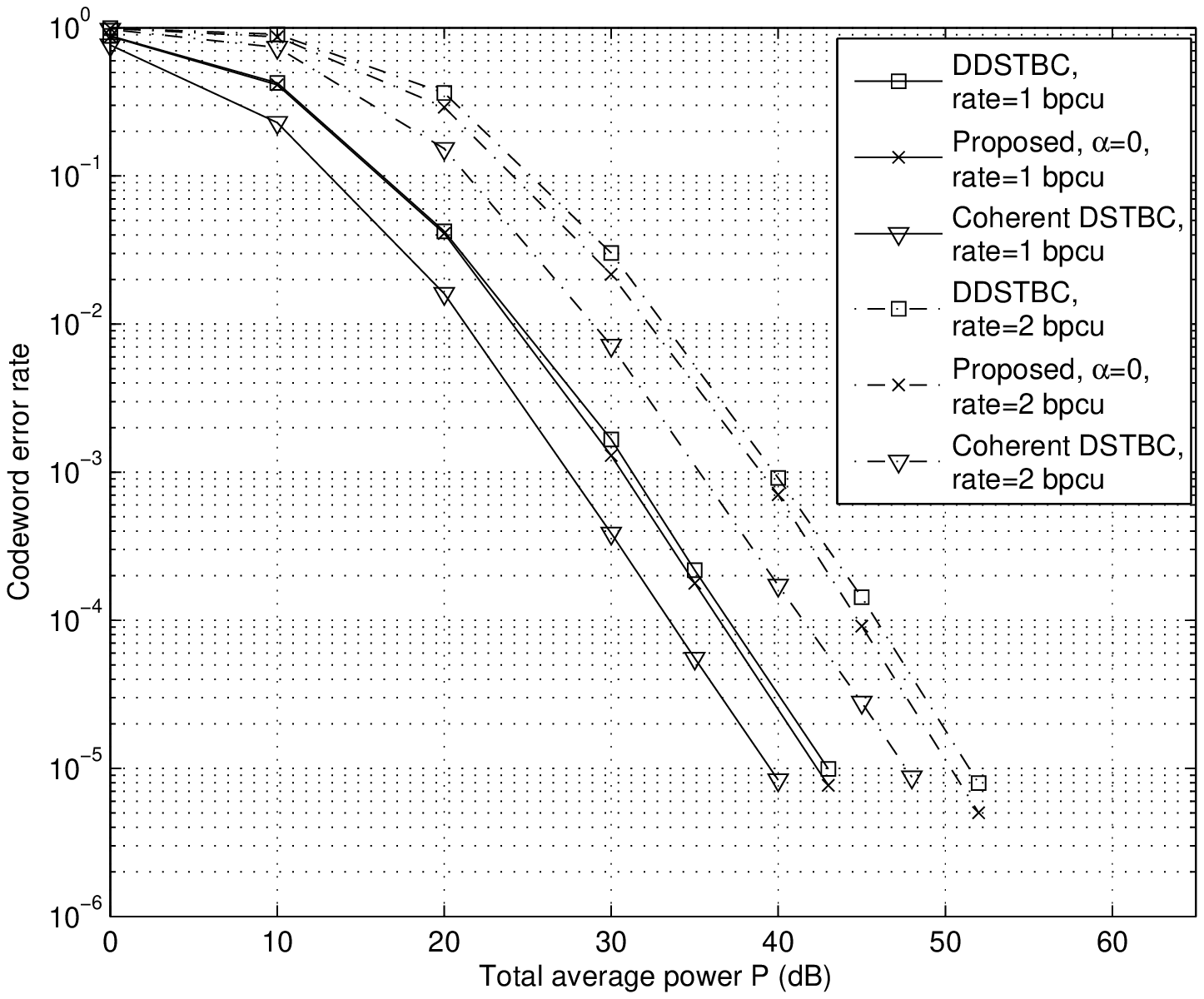}
\caption{Error performance comparison for a $2$ relay network}
\label{fig_simulation2}
\end{figure}

For a $2$ relay network, the Alamouti code is applied both as a DDSTBC\cite{JiJ} and as the underlying coherent STBC in the proposed training scheme. The signal constellation is chosen to be 4-QAM and 16-QAM for rates of $1$ and $2$ bpcu respectively. Fig. \ref{fig_simulation2} shows the error performance of the proposed strategy in comparison with Alamouti DDSTBC and the corresponding coherent DSTBC for $\alpha=0$ and transmission rates\footnote{When calculating transmission rate, the rate loss due to initial few channel uses for training is ignored ($R+1$ for proposed strategy and $2R$ for DDSTBC \cite{KiR,JiJ,OgH,RaR2}).} of $1$ bits per channel use (bpcu) and $2$ bpcu respectively. It can be observed that the proposed scheme has marginally better performance compared to the DDSTBC strategy for transmission rates of $1$ and $2$ bpcu. Note that the performance advantage of the proposed strategy over the DDSTBC strategy is more for the 2 bpcu case.

\begin{figure}[p]
\centering
\includegraphics[width=6.5 in]{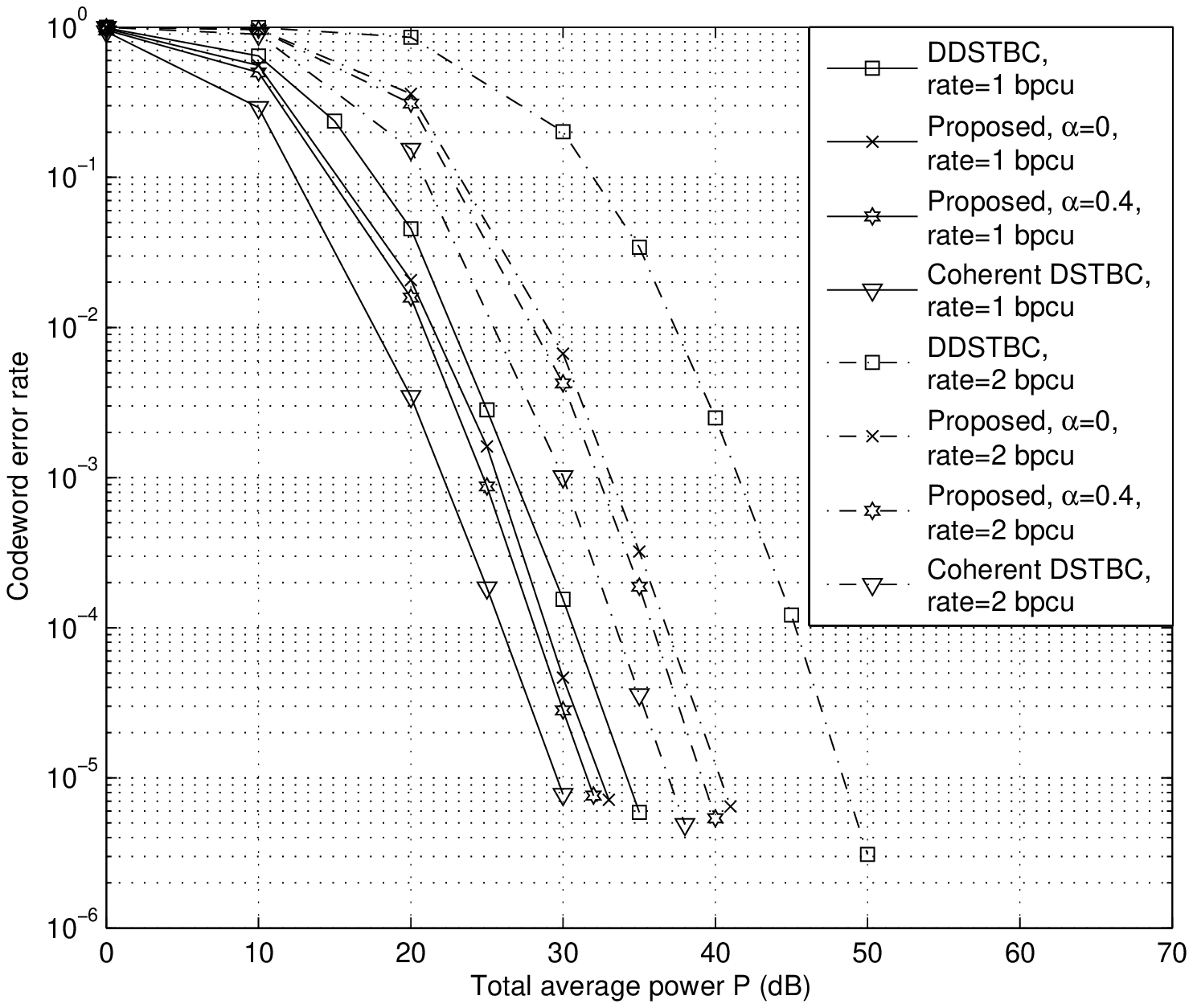}
\caption{Error performance comparison for a $4$ relay network}
\label{fig_simulation}
\end{figure}

For a $4$ relay network, the coherent DSTBC employed in the proposed strategy for simulations is $\left[\begin{array}{ccrr}
z_1 & z_2 & -z_3^* & -z_4^*\\
z_2 & z_1 & -z_4^* & -z_3^*\\
z_3 & z_4 & z_1^* & z_2^*\\
z_4 & z_3 & z_2^* & z_1^*
 \end{array}\right]$ where $\left\{\mathrm{Re}(z_1),\mathrm{Re}(z_2)\right\}$, $\left\{\mathrm{Re}(z_3),\mathrm{Re}(z_4)\right\}$, $\left\{\mathrm{Im}(z_1),\mathrm{Im}(z_2)\right\}$ and $\left\{\mathrm{Im}(z_3),\mathrm{Im}(z_4)\right\}$ take values from quadrature amplitude modulation (QAM) rotated by $166.7078^\circ$ (QAM constellation size chosen based on transmission rate). The relay matrices corresponding to this coherent DSTBC are unitary and $M=2$.  The DDSTBC taken for comparison is the one reported recently in \cite{RaR2}. It can be observed from Fig. \ref{fig_simulation} that for a rate of $1$ bpcu and codeword error rate (CER) of $10^{-5}$, the proposed strategy outperforms the DDSTBC of \cite{RaR2} by approximately $2$ dB for $\alpha=0$. For a transmission rate of $2$ bpcu, the performance gap between the proposed strategy and the DDSTBC of \cite{RaR2} increases to $8$ dB.  Finally, observe that a $40\%$ power boost to the pilot symbols gives marginally better performance (gain of $0.7$ dB). 

From all the above simulations, we infer that the performance advantage of the proposed strategy over DDSTBCs increases as the transmission rate increases. Also, note that the proposed strategy is better than the DDSTBCs of \cite{RaR2,JiJ} at all signal to noise ratio (SNR). In spite of the simple channel estimation method employed (Eq. \eqref{eqn_ch_estimator}), note from Fig.\ref{fig_simulation2} and Fig.\ref{fig_simulation} that the performance loss due to channel estimation errors is only about $3$ dB for transmission rates of $1$ and $2$ bpcu respectively. We can attribute three reasons for the proposed strategy to outperform DDSTBCs as follows: (1) lesser equivalent noise power seen by the destination during data transmission cycle as compared to distributed differential space time coding \cite{KiR,JiJ,OgH,RaR2}, (2) no restriction of coherent DSTBC codewords to unitary/scaled unitary matrices as is the case with DDSTBCs \cite{KiR,JiJ,OgH,RaR2} and (3) the relay matrices $\mathbf{B}_i,~i=1,2,\dots,R$ need not satisfy certain algebraic relations involving the codewords (see \cite{JiJ,RaR2} for exact relations), thus giving more room to optimize the minimum determinant of difference matrices (coding gain).

Simulation results are not reported for the asynchronous case because the use of OFDM essentially makes the signal model in every sub-carrier similar to the synchronous case. Except for a rate loss due to CP, the performance will thus be same as that for the synchronous case.

\section{Conclusion}
\label{sec5}

Similar to the results of \cite{DBV} for point to point MIMO systems, a simple training and channel estimation scheme combined with the protocol in \cite{JiH} was shown to outperform distributed differential space time coding at all SNR. The proposed strategy leverages existing coherent DSTBCs \cite{RaR_IT,KiR2,JiJ2,EOK,MaH} for noncoherent communication in AF relay networks. Finally, the proposed strategy is extended for application in asynchronous relay networks with no knowledge of the timing errors using OFDM. Some of the interesting directions for further work are: (1) design of optimal training sequences, (2) better channel estimation techniques and (3) optimal power allocation between the training cycle and the data transmission cycle.



\begin{figure}[p]
{\footnotesize
\begin{center}
\begin{tabular}{|c|c|c|c|c|c|c|c|}
\hline
Terminal & Slot $1$ & Slot $2$ & $\dots$ & Slot $M+1$ & Slot $M+2$ & $\dots$ & Slot $R+1$\\
\hline
Source & $\sqrt{\pi_1 P_t}$ & & & & & &\\
\hline
Relay $1$ & & $\sqrt{\frac{\pi_1\pi_2P_t^2R}{\pi_1P_t+1}}f_1$ & & & & &\\
& & $+\sqrt{\frac{\pi_2P_tR}{\pi_1P_t+1}}n_1$ & & & & &\\
\hline
\vdots & & & $\ddots$ & & & & \\
\hline
Relay $M$ & & & & $\sqrt{\frac{\pi_1\pi_2P_t^2R}{\pi_1P_t+1}}f_M$ & & &\\
 & & & & $+\sqrt{\frac{\pi_2P_tR}{\pi_1P_t+1}}n_M$ & & &\\
\hline
Relay $M+1$ & & & & & $\sqrt{\frac{\pi_1\pi_2P_t^2R}{\pi_1P_t+1}}f_{M+1}^*$ & & \\
& & & & & $+\sqrt{\frac{\pi_2P_tR}{\pi_1P_t+1}}n_{M+1}^*$ & &\\
\hline
\vdots & & & & & & $\ddots$ & \\
\hline
Relay $R$ & & & & & & & $\sqrt{\frac{\pi_1\pi_2P_t^2R}{\pi_1P_t+1}}f_{R}^*$\\
& & & & & & & $+\sqrt{\frac{\pi_2P_tR}{\pi_1P_t+1}}n_{R}^*$\\
\hline
\end{tabular}
\end{center}
}
\caption{Training cycle}
\label{fig_training}
\end{figure}

\begin{figure}[p]
{\footnotesize
\begin{center}
\begin{tabular}{|c|c|c|c|c|}
\hline
Terminal & \multicolumn{2}{c|}{Data transmission} &  & Data transmission\\
& \multicolumn{2}{c|}{cycle $1$} & $\dots$ & cycle $F$\\
\cline{2-5}
& Phase I & Phase II & &\\
& Slots $R+2$ & Slots $2R+2$ & & Slots $R(2F-1)+2$\\
& to $2R+1$ & to $3R+1$ & $\dots$ & to $R(2F+1)+1$\\
\hline
Source & $\sqrt{\pi_1P_d}\mathbf{z}$ & & &\\
\cline{1-3}
Relay $1$ & & $\sqrt{\frac{\pi_1\pi_2P_d^2}{\pi_1P_d+1}}f_1\mathbf{B_1}\mathbf{z}$ & &\\
& & $+\sqrt{\frac{\pi_2P_d}{\pi_1P_d+1}}\mathbf{B_1}\mathbf{v_1}$ & &\\
\cline{1-3}
\vdots & & \vdots & &\\
\cline{1-3}
Relay $M$ & & $\sqrt{\frac{\pi_1\pi_2P_d^2}{\pi_1P_d+1}}f_M\mathbf{B_M}\mathbf{z}$ & & \vdots\\
& & $+\sqrt{\frac{\pi_2P_d}{\pi_1P_d+1}}\mathbf{B_M}\mathbf{v_M}$ & &\\
\cline{1-3}
Relay $M+1$ & & $\sqrt{\frac{\pi_1\pi_2P_d^2}{\pi_1P_d+1}}f_{M+1}^*\mathbf{B_{M+1}}\mathbf{z}^*$ & &\\
& & $+\sqrt{\frac{\pi_2P_d}{\pi_1P_d+1}}\mathbf{B_{M+1}}\mathbf{v_{M+1}}^*$ & &\\
\cline{1-3}
\vdots & & \vdots & &\\
\cline{1-3}
Relay $R$ & & $\sqrt{\frac{\pi_1\pi_2P_d^2}{\pi_1P_d+1}}f_R^*\mathbf{B_R}\mathbf{z}^*$ & &\\
& & $+\sqrt{\frac{\pi_2P_d}{\pi_1P_d+1}}\mathbf{B_R}\mathbf{v_R}^*$ & &\\
\hline
\end{tabular}
\end{center}
}
\caption{Data transmission for the case $T_1=T_2=R$}
\label{fig_data}
\end{figure}

\begin{thebibliography}{1}

\bibitem{LaW} J.N. Laneman and G.W. Wornell, ``Distributed Space-Time Coded Protocols for Exploiting Cooperative Diversity in Wireless Networks,'' \emph{IEEE Trans. Inf. Theory}, vol. 49, no. 10, pp. 2415-2425, Oct. 2003.

\bibitem{JiH} Y. Jing and B. Hassibi, ``Distributed Space-Time Coding in Wireless Relay Networks,'' \emph{IEEE Trans. Wireless Commun.}, vol. 5, no. 12, pp. 3524-3536, Dec. 2006.

\bibitem{KiR} Kiran T. and B. Sundar Rajan, ``Partially-Coherent Distributed Space-Time Codes with Differential Encoder and Decoder,'' \emph{IEEE J. Select. Areas Commun.}, vol. 25, no. 2, pp. 426-433, Feb. 2007.

\bibitem{JiJ} Y. Jing and H. Jafarkhani, ``Distributed Differential Space-Time Coding for Wireless Relay Networks,'' \emph{IEEE Trans. Commun.}, vol. 56, no. 7, pp. 1092-1100, July 2008.

\bibitem{OgH} F. Oggier, B. Hassibi, ``Cyclic Distributed Space-Time Codes for Wireless Networks with no Channel Information,'' submitted to \emph{IEEE Trans. Inf. Theory}. Available online http://www.systems.caltech.edu/\~{}frederique/submitDSTCnoncoh.pdf.

\bibitem{RaR2} G. Susinder Rajan and B. Sundar Rajan, ``Algebraic Distributed Differential Space-Time Codes with Low Decoding Complexity,'' to appear in \emph{IEEE Trans. Wireless Commun.}. Available in arXiv: 0708.4407.

\bibitem{RaR_icc} ----, ``OFDM based Distributed Space Time Coding for Asynchronous Relay Networks,'' Proc. \emph{IEEE International Conference on Communications}, Beijing, China, May 19-23, 2008.

\bibitem{RaR_IT} ----, ``Multi-group ML Decodable Collocated and
Distributed Space Time Block Codes,'' submitted to \emph{IEEE Trans. Inf. Theory}. Available in arXiv: 0712.2384.

\bibitem{DBV} P. Dayal, M. Brehler and M.K. Varanasi, ``Leveraging Coherent Space-Time Codes for Noncoherent Communication Via Training,'' \emph{IEEE Trans. Inf. Theory}, vol. 50, no. 9, pp. 2058-2080, Sep. 2004.

\bibitem{GuX} X. Guo and X.-G. Xia, ``A Distributed Space-Time Coding in Asynchronous Wireless Relay Networks,'' \emph{IEEE Trans. Wireless Commun.}, vol. 7, no. 5, pp. 1812-1816, May 2008.

\bibitem{ZLiX} Z. Li and X.-G. Xia, ``A Simple Alamouti Space-Time Transmission Scheme for Asynchronous Cooperative Systems,'' \emph{IEEE Signal Processing Letters}, vol. 14, no. 11, pp. 804-807, Nov. 2007.

\bibitem{YLiX} Y. Li and X.-G. Xia, ``A Family of Distributed Space-Time Trellis Codes With Asynchronous Cooperative Diversity,'' \emph{IEEE Trans. Commun.}, vol. 55, no. 4, pp. 790-800, April 2007.

\bibitem{EK} P. Elia and P. V. Kumar, ``Constructions of Cooperative Diversity Schemes for Asynchronous Wireless Networks,'' Proc. \emph{IEEE International Symposium on Information Theory}, pp. 2724 - 2728, July 9-14, 2006.

\bibitem{EKJ} P. Elia, S. Kittipiyakul, and T. Javidi, ``Cooperative diversity schemes for asynchronous wireless networks,'' Wireless Personal Communications, vol. 43, no. 1, pp. 3-12, Oct. 2007.

\bibitem{KiR2} Kiran T. and B. Sundar Rajan, ``Distributed Space-Time Codes with Reduced Decoding Complexity,'' Proc. \emph{IEEE International Symposium on Information Theory}, Seattle, July 9-14, 2006, pp.542-546.

\bibitem{JiJ2} Y. Jing and H. Jafarkhani, ``Using Orthogonal and Quasi-Orthogonal Designs in Wireless Relay Networks,'' \emph{IEEE Trans. Inf. Theory}, vol. 53, no. 11, pp. 4106 - 4118, Nov. 2007.

\bibitem{EOK} P. Elia, F. Oggier and P. Vijay Kumar, ``Asymptotically Optimal Cooperative Wireless Networks with Reduced Signaling Complexity,'' \emph{IEEE J. Select. Areas Commun.}, vol. 25, no. 2, pp. 258-267, Feb. 2007.

\bibitem{MaH} B. Maham and A. Hjorungnes, ``Distributed GABBA Space-Time Codes in Amplify-and-Forward Cooperation,'' Proc. \emph{IEEE Information Theory Workshop}, Bergen, Norway, July 1-6, 2007, pp. 189-193.



\end{thebibliography}
\end{document}